# An Empirical Investigation of Pull Requests in Partially Distributed BizDevOps Teams

Viktoria Stray
*University of Oslo and
SINTEF*
Oslo and Trondheim, Norway
stray@ifi.uio.no

Nils Brede Moe
*SINTEF Digital
SINTEF*
Trondheim, Norway
nils.b.moe@sintef.no

Marius Mikalsen
*SINTEF and
Norwegian University of Science and Technology*
Trondheim, Norway
marius.mikalsen @sintef.no

Elin Hagen
*Department of Informatics
University of Oslo*
Oslo, Norway
elinhage@ifi.uio.no

*Abstract*—In globally distributed projects, virtual teams are often partially dispersed. One common setup occurs when several members from one company work with a large outsourcing vendor based in another country. Further, the introduction of the popular BizDevOps concept has increased the necessity to cooperate across departments and reduce the age-old disconnection between the business strategy and technical development. Establishing a good collaboration in partially distributed BizDevOps teams requires extensive collaboration and communication techniques. Nowadays, a common approach is to rely on collaboration through pull requests and frequent communication on Slack. To investigate barriers for pull requests in distributed teams, we examined an organization located in Scandinavia where cross-functional BizDevOps teams collaborated with off-site team members in India. Data were collected by conducting 14 interviews, observing 23 entire days with the team, and observing 37 meetings. We found that the pull-request approach worked very well locally but not across sites. We found barriers such as domain complexity, different agile processes (timeboxed vs. flow-based development), and employee turnover. Using an intellectual capital lens on our findings, we discuss barriers and positive and negative effects on the success of the pull-request approach.

*Keywords—pull request, communication, coordination, distributed global teams, Slack, agile software development, human aspects, large scale, BizDev, case study, empirical research*

## I. INTRODUCTION

Offshore outsourcing (sub-contracting to third-party vendors from other countries) is one major trend within global software engineering (GSE). Outsourcing software development is most often motivated by developers' low hourly rates in other countries [1]–[3]. The lack of knowledge or resources is the second biggest motivational factor. In the increasingly popular IT field, jobs are abundant but developers who can fulfill those jobs are few. However, acquiring the right resources can be not only challenging but costly as well. Developing software in globally distributed teams is challenging because of cultural differences, language barriers, communication issues, and time differences [4]. Further, when work is outsourced, the challenges become even bigger [5].

To mitigate the challenges with knowledge sharing and coordination in GSE [6], [7], most companies have transferred to agile software development and cross-functional teams.

Applying agile to a distributed setting affects how virtual teams are structured and work together, relying much on teamwork. In a distributed agile setting, a common strategy for coordination across sites involves meetings using online collaboration tools such as Slack [8]. While agile methods, meetings, and tools reduce alignment problems between development teams and management, alignment problems between business, development, and operations in GSE still exist. Therefore organizations create DevOps [9], or BizDevOps [10], [11], teams. When software development and IT operations are aligned, it is called DevOps [12]. Gruhn and Schäfer [10] explain the term BizDevOps as follows: "Business, Development and Operations work together in software development and operations, creating a consistent responsibility from business over development to operations." BizDevOps teams consist of people from various organizational functions (e.g., enterprise architecture, business development, software development, testing, and operations). However, extensive cultural training is needed for such teams to succeed [13].

Previous studies show that the main reason that companies terminate offshore contracts is the low quality of software being developed [5]. Extensive testing is one approach to ensuring quality in distributed software development [14], [15]. Another approach is using software code review [16]. Although the code review process was previously time consuming, the practice has evolved along with the industry and is now incremental and lightweight [17]. The formal software inspection process has been replaced by a mechanism called a *pull request* (PR), a term introduced by GitHub [17]–[20]. While the PR mechanism is now standard for distributed code reviews, challenges have nonetheless been reported, and the practice has been found to both increase and decrease the speed of the code review process [21]. Code reviews do more than just ensure quality. Because developers collaborate through code during a review [22], such reviews enable knowledge sharing and aim to balance the skills in the teams [23], which is important in distributed teams.

Even though new techniques, processes, and approaches to GSE are likely to increase quality, Moe et al. [5] found that the main reason for the quality problems in sourcing relationships was not being able to build necessary human and social capital (i.e., individual creativity and the relationships between team members). Indicative of this are the reported challenges with domain knowledge and high turnover, which have only amplified the GSE problems. Even if companies introduce new techniques and approaches such as BizDevOps and PRs, it is unlikely they will solve all problems related to collaborations in an outsourcing relationship. In this study, we therefore asked the following research question:

*What factors affect a PR approach in distributed agile BizDevOps teams?*

To reach an answer, we report findings from an interpretative case study of a large agile program, consisting of BizDevOps teams that were partially distributed with parts of the outsourced team located offshore. A qualitative research method was chosen as the best way to "focus on discovering and understanding the experiences, perspectives, and thoughts of participants – that is, qualitative research explores meaning, purpose or reality" [24].

The rest of this article is organized as follows. In Section II, we provide the theoretical background, discussing PRs and intellectual capital in GSE. Section III details our data collection and analysis, and provides details on the interpretative case study method. Section IV presents findings related to the large-scale distributed context and barriers to using PRs in the distributed setting. Section V discusses these findings in light of the concept of intellectual capital. Section VI provides implications of our research, and Section VII offers conclusions and suggestions for future work.

## II. Background

To save development costs, offshoring development work to another country is one of the key strategies companies use, often due to the low hourly developer rates [1]. However, outsourcing has many extra or hidden costs that can lead to it not being a money-saving initiative as anticipated [3], [25], [26]. Additionally, the need for resources and competence is also a motivational factor for outsourcing work. In this section, to understand PR in the context of GSE, we present a background of PR research and intellectual capital.

### A. Pull requests in global software development

The success of GSE greatly depends on effective knowledge sharing within and between software development teams [6], [7]. A lack of knowledge sharing has been reported as one of the main challenges in GSE [27]. Further, effective knowledge sharing helps distributed teams collaborate more effectively {Citation}. Sharing knowledge within a team that is globally distributed is very difficult, and therefore tools for quality assurance (QA) of code are put in place [28]. PRs are vital for QA in the software development workflow and have become the standard mechanism for reviewing distributed code [21]. A key benefit of a PR approach is that the technique facilitates knowledge sharing, is highly collaborative, and is a lightweight, modern QA mechanism [17]. In the PR approach, a contributor creates a PR after making code changes, and then a reviewer inspects the suggested changes to see whether they can be merged into the project. The reviewer then interacts with the contributor and others in discussion threads associated with the PR [21]. In global software development, trust has been found to be a key factor for accepting PRs [19].

PRs thus have the potential to support the challenge of knowledge sharing in GSE by being a new way of reviewing code in distributed teams. Code reviews can be viewed as positive for mentoring and seen as opportunities to shape the codebase [22]. Other benefits of code reviews are as follow [23]:

1. Better code quality because knowing that someone will review the code has a preventive effect
2. Fewer defects
3. Reviewers learn because they receive knowledge about the changed code and how to solve problems
4. Authors learn from receiving feedback on the code they wrote, and they learn about possible new solutions, new libraries, and the reviewers' values and quality norms
5. Sense of mutual responsibility and collective code ownership
6. Better solutions
7. Complying with QA guidelines.

Although many benefits stem from code reviews, PRs may slow down the software development process when the team members do not actively engage with the PRs and review them in a timely manner [21]. For example, when the developers doing the review are overloaded with other tasks or prioritize other things, then a PR can remain open for a long time, slowing down the overall coding process and even causing merge conflicts. The size of the PR also greatly affects completion; developers prefer smaller PRs over larger ones [29]. Other undesired effects of reviewing PRs is that it demands more staff, the cycle time increases, and the reviewer might offend or discourage the PR author [23]. Paul et al. [30] recently found that males are more negative and less encouraging when giving comments in code reviews to females than they are to other males, and negative comments in PRs may demotivate developers. The above points to successful PR being dependent upon how the code review work is organized as knowledge-intensive work. Next, we explore intellectual capital.

### B. Intellectual capital

Wohlin et al. [31] proposed a general theory of software engineering for balancing human, social, and organizational capital and suggested that these three components make up the sum of the intellectual capital within an organization. Intellectual capital is a particularly relevant perspective for software development because software development is knowledge-intensive work. Moe et al. [5] found that the main reason for terminating sourcing relationships is the inability to build the necessary human, social, and organizational capital. Human capital is the skills and knowledge leading individuals to provide solutions. This capital "resides with, and is utilised by individuals" [31]. An individual's creativity also falls under this category. Social capital consists of knowledge resources embedded within, available through, and derived from a network of relationships [32]. Such relationships are not limited to internal knowledge exchanges among team members, but extend to linkages with customers, suppliers, alliance partners, and similar [5]. Organizational capital can be defined as the "possessions remaining in the organization when people go home" [31]. This largely concerns source code, processes, various documentation, culture, and infrastructure. Intellectual capital is all the knowledge in a company and consists of the three intellectual capital components: human, social, and organizational knowledge [31]. These three components should be balanced to sufficiently carry out a task in a cost-efficient way in distributed development [31]. When intellectual capital is too low, tasks will not be implemented sufficiently, and when intellectual capital in a sourcing relationship is too high, it will probably lead to higher costs.

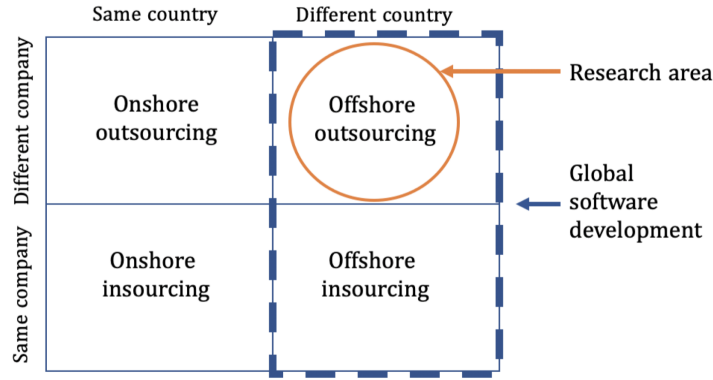

Fig. 1. Types of sourcing arrangements and our research area, adapted from [2].

Finding the right level of intellectual capital depends on the objective of the task (i.e., the intended level of performance). A developer with little experience will probably rely on good source code and software documentation (organizational capital), as well as the expertise of others (social capital), to fulfill the objective of the task [31]. On the other hand, the same level of high organizational capital may be redundant for an experienced developer who has good knowledge of the code. Social capital is also interesting to examine because it might encourage the development of intellectual capital. For example, good collaboration with other external units or experts may lead to individual learning, thus increasing the human capital, which again increases the intellectual capital.

To handle PR tasks in a global context, developers need to follow and understand the process, collaborate with team members, trust them, and understand the domain and programming knowledge. In sum, developers need the right level of intellectual capital.

III. CASE DESCRIPTION AND METHOD

A. Case study design and context

This study is an interpretative case study [33] of a large-scale program in which we report on barriers to distributed developments and the use of PRs. We have closely followed the case since 2014. The case study was conducted in a large Nordic company called NorBank (pseudonym). NorBank has 2000 employees and in-house software development units in Sweden and Norway.

Smite and Wohlin [2] present different types of global software engineering, concentrating on off-, on-, near-, and far-shoring, as well as insourcing and outsourcing. The context of the present study is offshore outsourcing, which means work is performed in a different country (offshore) by an external third-party collaborator (outsourcing) [26] (Figure 1).

Previously, NorBank had a long-term relationship with a Ukrainian subcontractor, which was terminated. In 2016, NorBank established a new relationship with a large consultancy company in India, which we named KappaTech (pseudonym) in this paper. Smite and Wohlin [2] described sourcing as a form of collaboration, whereas Oshri et al. [34] define sourcing as "the act through which work is contracted or delegated to an external or internal entity that could be physically located anywhere." The two companies considered the collaboration a partnership. At the beginning of the sourcing relationship, all KappaTech personnel were moved onshore for several weeks. Some stayed for months, and the motivation was to speed up the onboarding process and create a strong relationship between the developers in the two companies.

TABLE I. DESCRIPTION OF ROLES IN THE BIZDEVOPS TEAMS

| Role | Responsibilities |
|---|---|
| *Developer* | Included both back-end and front-end developers, with juniors, seniors, tech lead, and domain architect. Performed development, reviewed code, and assigned which tasks should be developed at which site. |
| *Test leader* | An administrative function that made sure the deliveries were thoroughly tested. The test leader had an overview of the tasks in Jira and made sure they had the testers needed. |
| *Business developer* | The business developers contributed to planning the development, limited work by designing minimum viable products, created Jira tasks, and coordinated with business and other departments. The business developers also conducted user testing of the product. |
| *Product owner* | The product owner had overall responsibility for reaching targets and for the teamwork. The product owner made a development plan along with the enterprise architect, removed obstacles, kept stakeholders updated on the status of their projects, their plans, and upcoming deliveries. The product owner also acted as a buffer when requests were made to the team and coordinated work with the other teams in the large-scale setup. |
| *UX designer* | The UX designer was administrative for all the UX work and made sure the company had a uniform profile. |
| *Enterprise architect* | The enterprise architect had overall responsibility for the solution the team was making. The person was responsible for making sure the solution adhered to the rest of the enterprise's solutions and that the system the team was building reflected the strategies of the department. The enterprise architect also worked on coordinating the platform with others. |
| *Data scientist* | The data scientist was responsible for implementing Google Analytics tracking, helping business developers analyze and understand the data collected, and helping with making decisions based on the data. |

The teams we studied in Norbank were organized as BizDevOps teams. In such cross-functional teams, representatives from former business, IT, and operations work together. Descriptions of the different roles in the BizDevops teams in Norbank are shown in Table I. The large-scale setup with the five teams is described in more detail in Section IV.

*B. Data Collection and Analysis*

Data were collected between September 2018 and April 2019. We conducted 14 interviews, observed 23 entire days with the team, and observed 37 meetings. Additionally, we collected various documents.

In advance of the interviews, each participant was given the prefix N and then a random number. The document linking the informants' names and numbers was password protected to ensure total anonymity. All of the interviews were recorded, transcribed, and saved together with the randomized participation number.

We interviewed members from two different teams: Alpha and Kappa 1. Two representatives were from Kappa 1 located on-site and 11 persons were from team Alpha. In addition, we interviewed the offshore delivery coordinator who was responsible for coordinating the activity between the Nordic teams and the KappaTech teams. The interviews with the two representatives from Kappa 1 were held in December 2018 and transcribed within a few days. The other interviews were conducted over two weeks in January 2019 and transcribed within a week of the final interview. The interviews lasted from 26 minutes up to 1 hour, with an average of 46.7 minutes. Table II gives an overview of the different roles of the people interviewed.

We also observed how people located on-site were working and we observed 37 meetings, as shown in Table III. Another source of evidence was documentation. We collected pictures, documents, presentations, and reports. These included the teams' presentations and progress plans, as well as analysis results from surveys. The documentation was helpful to gain a better understanding of the context. Additionally, documentation was useful for verifying specific details.

TABLE II. OVERVIEW OF THE INTERVIEWS

| Role | No. of persons interviewed | Team | Duration in minutes |
|---|---|---|---|
| Developer | 1 | Kappa 1 | 48 |
| Point of contact | 1 | Kappa 1 | 53 |
| Product Owner | 1 | Alpha | 59 |
| Developer | 4 | Alpha | 42, 42, 47, & 51 |
| Test Leader | 1 | Alpha | 49 |
| Business developer | 2 | Alpha | 41 & 60 |
| Enterprise architect | 1 | Alpha | 51 |
| UX-designer | 1 | Alpha | 37 |
| Data Scientist | 1 | Alpha | 47 |
| Offshore Delivery Coordinator | 1 | Other | 27 |
| Total | 14 | | Average: 46.7 |

TABLE III. OVERVIEW OF OBSERVED MEETINGS

| Type of meeting | Number of observed meetings | Average number of participants | Average duration in minutes |
|---|---|---|---|
| *Stand-up with Alpha and Kappa 1* | 6 | 10.5 | 10 |
| *Weekly stand-up* | 5 | 10.3 | 25 |
| *Weekly progress meetings* | 4 | 4 | 52 |
| *Team workshops* | 2 | 12.5 | 60 |
| *Team retrospectives* | 2 | 8.5 | 60 |
| *Project retrospectives* | 2 | 11 | 105 |
| *Team-related meetings* | 10 | N/A | N/A |
| *Project-related meetings* | 6 | N/A | N/A |

The data analysis was conducted in four main steps. First, we started collecting the data before deciding on which theory to use. Before the two first interviews, we had categorized the questions in certain overall topics, such as processes and teamwork. Every observation was documented, and for the greater part of the observations, a reflection note was written containing initial reflections on what was occurring. Second, transcribed data were entered into the NVivo qualitative analysis software. Text was coded into specific nodes, and then nodes were categorized into the different aspects of intellectual capital. Third, after identifying code quality as a concern in the case, we coded for specific positive and negative experiences with PRs. Twenty-seven codes were generated during focused coding (e.g., "QA meetings," "PRs perceived as tedious," and "Use of Slack").

IV. FINDINGS

In this section, we first present the background and context of the studied case, the way they were set up, and their development process. Then, we describe the developers' experience of reviewing PRs in this partially distributed context.

*A. Large-scale distributed context*

In 2017, NorBank initiated an agile program that consisted of four cross-functional autonomous teams organized in line with agile principles to develop software for their business-to-business solutions in the insurance market. The teams consisted of resources from both the software and business development sides of the organization. The teams delivered software solutions (e.g., sales and settlements) to the business side of the organization. The teams collaborated closely with organizational units responsible for technology development and innovation. Product managers who were part of the program's steering-forum and managers from the business and technology units led each team. We named the agile program Terra. Alpha was the team we followed closely. The Beta, Gamma, and Delta teams were mainly observed through meetings. Alpha and Beta were extended with teams from KappaTech (Kappa 1 and Kappa 2; Figure 2).

Team Alpha consisted of 13 people when first starting the study, including members with part-time positions in the team.

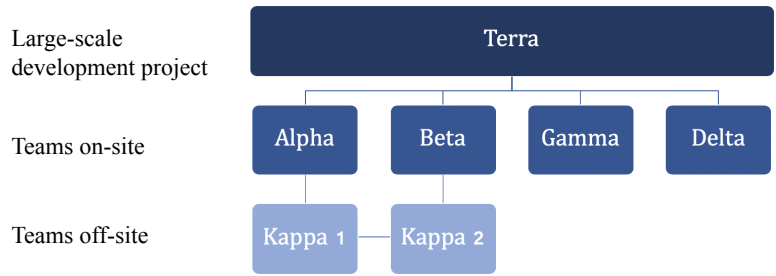

Fig. 2. Large-scale project structure

There were six developers, one test leader, two business developers, one product owner, one UX designer, one enterprise architect, and one data scientist. Not only was team Alpha a part of the large-scale Terra project and cooperating with KappaTech teams, it also had six other teams and departments in the organization with which it coordinated and, as such, had many dependencies the team had to manage. The Alpha team was described as a BizDevOps team (see Table I for the description of the different roles in this team).

The teams in KappaTech had 23 members. Although the Norbank mostly had members dedicated to the teams, members sometimes also shifted around depending on where the workload pressure was high. In KappaTech, there was one project manager, one architect, one automation developer, two team leads, 14 developers, and four testers.

Team members from Alpha and Kappa 1 had visited each other on several occasions, and two people from Kappa 1 had been placed on-site site between mid-2016 and October 2018. This had helped a great deal with the collaboration between the sites, as illustrated by N3: "*The [point of contact from KappaTech] works well. He is a very skilled guy and has control over a lot. He has understood much of the business and can function as a good interpreter between NorBank and KappaTech.*"

N8, a developer, described having two members from KappaTech on-site was also described as "*a small revelation for us.*" N11 (the test leader) said, "*Problems were solved much quicker.*" However, this setup also made much of the communication go through the two persons from KappaTech because it was easier.

Alpha was responsible for managing the progress plan and decided which development tasks should be developed at which site. Alpha team members also wrote the product specifications for the tasks. They usually did not collaborate on tasks across sites.

Several tools were used to coordinate the development tasks. By using Jira, all team members, developers in particular, had the possibility of keeping track of who was solving which tasks. Most interviewees stated that they had a good overview of what other people were doing, though not necessarily always in detail.

Slack was another tool that gave a good overview of what was occurring. Slack was used for most of the communication and it was important for their agile process. They also used Slack for many of the daily stand-up meetings. N8 said, "*It works much better with a small stand-up in Slack so people know what you are doing. There is no point in standing in a circle and wasting time.*" N10 described both positive and negative sides of using Slack:

> *The flow of information on Slack, I am very happy with that. The amount of emails has drastically gone down, thank goodness. I used to spend an awful lot of time on e-mails. The only negative with the use of Slack is that it is hard to keep the work-life balance and quit working; you are on work all the time. It is probably the toughest challenge with that type of communication, but I think both the tone we have on Slack and team-wise it is very good!*

*B. Pull requests in agile distributed teams*

The developers in team Alpha reviewed each other's PRs. In addition, three of the on-site developers reviewed the PRs sent from Kappa 1. To reduce rework and increase the PRs' probability of being accepted, Kappa 1 did an internal code review before submitting its PR. While this process, in theory, should work fine, we found several frustrated team members. We identified four repeating topics when interviewing them about the PR process, which we will describe next.

*1) Domain complexity*

Because of European bank regulations and policies, the project was set up in such a way that the on-site developers were responsible for the quality of all the code developed. In practice, it meant an on-site developer had to approve all code developed off-site. Alpha team members rejected many of the PRs. When compared to the other teams working on Terra, we found the rejection number to be much higher in Alpha regarding PRs from KappaTech than it was for the other on-site teams, even though they worked with the same pool of developers. Even though we did not investigate the exact reason for the rejection of PRs, in a meeting, the person responsible for the collaboration between the sites commented that the domain was much more complicated for the Alpha team than it was for the Beta, Gamma, and Delta teams.

The domain becoming a challenge was surprising because KappaTech had experience from similar projects in the same domain. When interviewing a developer from Kappa 1 (N4) located on-site about the role of domain knowledge, he explained that "*it's not just about the business model, it's about the insurance policy. The way insurance works here is not a bit different—it's a lot different than in [the homeland] and other countries we have worked on.*" Because the on-site KappaTech developer worked closely with the on-site team and spent 60% of his time coding, he acquired more experienced over time regarding the domain. However, it became evident that the off-site developers improved slowly. N14 continued: "*We need to figure out more ways to communicate this kind of business knowledge ... so everything gets notified and gets broadcasted to everybody that is in the team so that it's just not depending on one person.*"

*2) Agile methodology and work habits*

Both teams used agile methods, and they had the freedom to choose their processes, practices, and tools. Alpha, the on-site team, used Kanban, which is a flow-based approach. To reduce interruptions, team Alpha tried to hold meetings only on Tuesdays and Thursdays to enable them to concentrate for full days the rest of the week. Consequently, some of the stand-up meetings were conducted on Slack. Team Kappa 1 followed Scrum, which is a timeboxed approach, and the team worked in two-week sprints. One reason was to have more control, as described by N1:

*With KappaTech, we need more control and to be more rigid, so there we run three-week cycles (week zero, one and two). That is, we have two-week sprints with preparations before, and also demos and retrospectives. So, with them we have a quite well-defined set of ceremonies that we run.*

Using Scrum was seen as beneficial for KappaTech because shielded the Nordic developers from having to review PRs daily but instead do it more intensely over one week. While reviewing others' code was a key activity, it was not seen as the most exciting work to do, independent of the code being written on-site or off-site.

Because of KappaTech's decision to follow Scrum, after each sprint, a large number of PRs was sent to Alpha for review. It was a bit overwhelming for the Alpha developers receiving the PRs, as illustrated by N3, a developer:

*When I get QA from KappaTech, then it is maybe 64 [pull requests], so then it's like, 'Wow! Where should I begin?' That is cumbersome, not very motivating. Reviewing pull requests is never particularly fun regardless of who sent them, and now it has been quite a lot. Then it gets really boring.*

Further, participants said the off-site PRs were more extensive (because they had been coding for weeks) than the on-site PRs, which made the work even more demotivating because an extensive PR takes more time and energy to approve as it becomes more complicated.

The amount of PR work stayed high over the whole period and did not change much. Multiple on-site team members were worried that some of the senior developers might quit because of this tedious task. N1 explained: "*I'm scared that some of these, like [name of developer], who is an incredibly competent and experienced developer and been here for a very long time, will quit because he can't take it anymore.*" N1 also said that the part of his job that made him most unhappy and frustrated was when he had to review PRs from Kappa 1. While the seniors were unhappy, one on-site junior developer explained that he learned while reviewing code, so he did not mind doing it.

*3) Social networks*

Reviewing PRs was seen as boring work. While large PRs and many PRs arriving at the same time seemed to reduce job motivation, an additional factor seemed to influence motivation positively; that is, knowing the person who had written the code. In addition, reviewing PRs for the on-site developers was more satisfying because they could sit together. N3 explained: "*If we do QA locally, then we do it together with the developer, who is also someone I know. So, I think it's easier and then you can talk back and forth, so I think it's much better*". N12 elaborated on why he found it easier to do pull requests from people he knew: "*[Name of developer] and I have worked here for a long time and we know each other, we trust each other. I know that most of what he does is of high quality.*"

Not all Alpha and Kappa 1 team members had met physically, and consequently not everyone knew each other. Alpha members stated multiple times that they were unsure about the knowledge of each of the Kappa 1 members. N1 said, "*We don't really know their competence properly.*" Further, KappaTech sometimes added new people to the Kappa teams and a high attrition rate (turnover) increased the problem of working together on PRs. Based on a resource overview from KappaTech, we analyzed the attrition numbers. Of the 29 people at KappaTech that joined Terra in August 2016, after 2 years, 16 people had left and 12 had join. The annual turnover rates were calculated to be the following:

- 2016:  4%
- 2017: 18%
- 2018: 38%

We were told that most of the new people were junior developers. N8 explained the effect of having many new people on the project:

*I feel there are too often new names appearing in the pull requests, names I have never seen before.... When you know the person, you know his strengths and weaknesses, and then it becomes much easier to press 'approve' because I know what he knows. Now, it's more like I have to analyze the code much more carefully, because I do not know who wrote it and how good an understanding that person has of our complex domain.*

N13 described in detail how the communication on Slack let everyone know what was occurring in the distributed project. Because the communication was more informal than the communication via e-mail was, the threshold for communicating was lowered across sites, and team members could quickly clarify PRs. Further, the structure at KappaTech was described as more hierarchical and traditional than at Norbank, thus making it harder to rely on informal and frequent communication, but the use of Slack helped the situation. N5 said, "*People are not afraid to write both small and big things on Slack.*"

Although they were using Slack for clarifications, they needed to change the collaboration process to improve the situation. Therefore, they introduced additional quality meetings where they looked at the PRs. N15 from Kappa 1 explained how he was pleased that these meetings were introduced because they made the collaboration more effective:

*If someone sits and tries to think of why we did something, [then] that takes a lot of time. So it's better that we show them this is why we did what we did, and explain the thought process behind the code.*

*4) Measuring productivity*

The KappaTech team was frequently measured by key performance indicators (KPIs) set by NorBank. The KPIs were measurements from numbers they received from queries in Jira. The queries retrieved information such as how long a task spent at a certain stage in the development process and how many tasks team Alpha rejected and sent back to KappaTech. Although many PRs were rejected and the quality was not seen as satisfactory, nevertheless the KPIs on the code and productivity were met. Moreover, as long as the KPIs were met, managers at KappaTech and Norbank were happy. When investigating why the KPIs on the PRs were shown as good despite the quality issues, we found two explanations. The KPIs did not measure the acceptance rate on GitHub, but rather the acceptance rate in Jira. Further, a PR could have many Jira tasks connected to it, meaning that when a PR was finally accepted, many Jira tasks were then also accepted. Moreover, when a Jira task was split into many sub-tasks and they were all approved, it looked as though much work had been done and that the quality was good. Consequently, the acceptance rate KPI was good. N8 explained:

> *The KappaTech teams are measured by the number of pull requests that are approved or not, but for some reason that measurement is calculated by completed Jira tasks and not actual PRs in the system where we approve the code. Often, I feel that one pull request has 10 attached Jira issues, to double the KPI. I think measuring an outsourcing partnership based on the number of approved PRs is not a good solution.*

## V. Discussion

Typical reasons companies terminate offshore contracts are the low quality of the software being developed and a knowledge gap between off-site and on-site workers [5]. Relying on PRs has been found to mitigate such barriers because the approach results in better code quality and facilitates knowledge sharing between team members [23]. However, PRs may also slow down the software development process when the requests are not reviewed promptly [21]. We studied the PR approach in an offshore outsourcing relationship using interviews and observations. We will next discuss the factors that affect the PR approach in distributed agile BizDevOps teams. We investigated the positive and negative factors that affected the PR approach and the ways they relate to the three components of intellectual capital [31], as shown in Figure 3.

*A. Factors that negatively affected the pull request approach*

The developers categorized performing QA, or code review, as one of the most boring tasks they had to do. We found that only a subset of the on-site team members acted as reviewers, making the number of tedious tasks high. Some people even threatened to quit if they had to spend too much time reviewing PRs. Even though the idea of two-week sprints and fewer interruptions was good on paper, it made the situation worse in practice. After each sprint, the developers received a very high number of PRs, and some were very large, which made the job more demanding and even less motivating. Periods with too many and too large PRs resulted in demotivated and frustrated developers. Our findings support other research that has shown developers prefer smaller PRs over larger ones [29], [35]. Another explanation for the on-site developers being dissatisfied is that team members who receive others' task outputs are less satisfied [36].

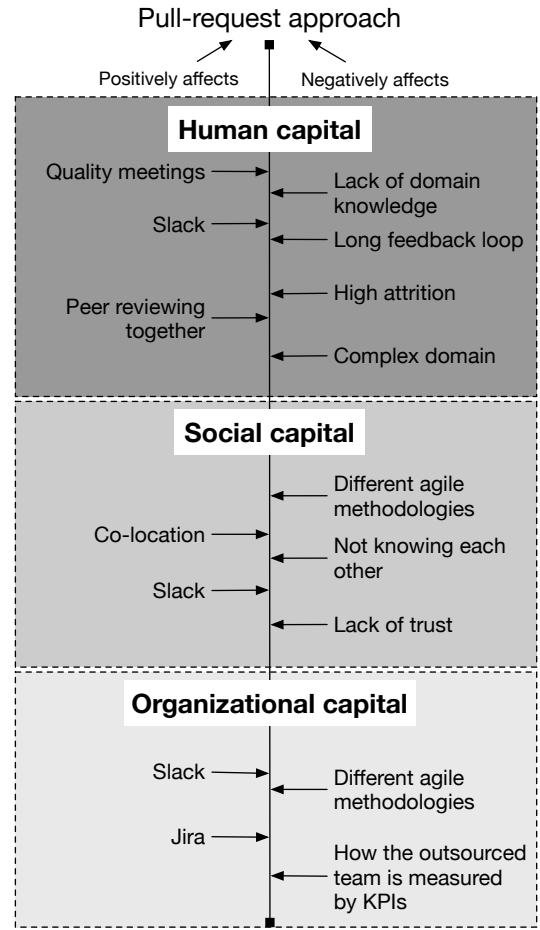

Fig 3. The factors that affected the pull request approach

The challenge caused by relying on two types of agile processes between distributed teams can be seen as a problem with the organizational capital [31]. While partially incompatible processes cause problems, the major challenge seemed to stem from the lack of human capital. The offshore developers lacked domain knowledge, which led to many PRs from the off-site developers being rejected. In another NorBank team, the rejection number was lower even though the PRs partly came from the same pool of developers, indicating that Alpha gave Kappa 1 team complex tasks. Our finding is in line with Wohlin et al. [31], who found that assigning simple and non-critical tasks and more minor product improvements when remote offshore developers climb the learning curve is essential. Similarly, Moe et al. [5] studied four outsourcing attempts and found high turnover and lack of domain knowledge affected the level of human capital in all cases.

The situation became worse because of the high attrition rate and people not knowing each other. Our findings are in accordance with Smite and van Solingen [26], who found high attrition increases the length of a remote team's learning curve. Not knowing each other resulted in developers not trusting the code submitted for review, as also found in [19], [37]. The low trust resulted in the Alpha members being very thorough when reviewing PRs from Kappa 1 because they did not know what human capital the Kappa 1 members possessed.

Social capital has been found to play an important role in PR processes. Social capital is the team members' network, and the organization supports the creation of social capital when it brings its members together to do their primary tasks and coordinate work, particularly in the context requiring mutual adjustment [31]. In open source projects, strong social connections increase the chances that a PR is accepted, and a weak connection reduces the chance of a PR being accepted [35]. Similarly, our case shows that a weak social network and weak social capital negatively affected the PR approach.

*B. Factors that positively affected the pull request approach*

While most developers perceived the work of doing PRs as a less motivating part of the job and something they did not want to do, they continually improved the practice. We found several factors that made the practice work better and become less tedious.

The on-site Alpha team members often worked in pairs when doing QA and development tasks. However, they did not pair up with the off-site developers, partly because of the time zone differences. Pairing with Kappa 1 might have enhanced their human capital, thus making Kappa 1 perform better later. Although they did not pair up across sites, they introduced joint quality meetings where they discussed the PRs. Another positive practice was that they had two people from KappaTech co-located on-site, which increased the on-site KappaTech employees' social capital and was found beneficial for the work. Our findings are in accordance with Smite et al. [38], who indicate that social capital and networking are essential when solving complex, unfamiliar, or interdependent tasks.

Coordination tools were also a positive factor for the PR practice. Slack has been found to support problem solving and knowledge sharing in distributed teams [8]. Our findings show that Slack supported the PR approach because developers discussed and shared knowledge (human capital) in the tool. However, a common challenge is that the knowledge documented in Slack logs, which is a form of organizational capital, [31] can be difficult to retrieve at a later time because of a high number of tools used (e.g., Slack, Jira, and GitHub), and developers might not remember where they discussed an issue [39]. Further, when new people join, the information stored in Slack might be hard to find.

*C. Implications for practice*

Our study generates several findings with practical implications. First, there is a need to focus on creating good agile teams across locations. Software development and handling PRs is teamwork that requires much communication and knowledge sharing. Thus, to ensure high quality and a good PR process, focusing on working as one team seems reasonable. While offshore outsourcing has a buyer–vendor aspect, autonomy and agility need to be in focus, thus ensuring more efficient collaboration.

Second, being attentive to employee attrition levels is important. High employee turnover is likely to happen in a sourcing relationship, and easy routine tasks are demotivating for the on-site personnel and may increase the attrition rate on-site. Establishing the right processes to maintain intellectual capital and balance the workload may help deal with high turnover. When a person leaves the team, his or her human capital should be preserved in new human capital, social capital, or organizational capital. Regulating turnover in the contract can also be a solution for dealing with this. In a study from 2016, Smite and Van Solingen [26] attempted to calculate offshore-outsourcing costs between a Dutch software company and an Indian vendor. They found that learning costs due to offshore employee turnover were the primary cost factor to control. In rapid-growth markets, high turnover rates are not unusual [5], [26]. Balancing the workload is key. If there is a need for the on-site team to handle all the PRs, then there cannot be too many remote developers. The correct ratio depends on the skills of the off-site developers (human capital), the network of the developer (social capital), and the development process in use (organizational capital). The importance of the intellectual capital components depends on the specific task. Therefore, a junior developer who has low human capital and lacks a network (social capital) will need to strengthen his or her social capital by building the network, simultaneously having adequate organizational capital as support. Human capital strengthens over time (e.g., by working with project tasks).

Third, PRs must not be too large. Our findings indicate that the amount of reviewing and rework resulted in a higher cost to the sourcing relationship and lowered work satisfaction on-site. Therefore, we suggest that companies aim to have their project members submit small PRs and follow technical contribution norms, thereby increasing the likelihood of the PR being accepted [35].

Finally, there is no off-the-shelf solution for making collaboration successful in an outsourcing setup; we found repeated issues pertaining to quality and motivational problems and indications of increased costs. Our findings align with Smite et al. [1], who argued that outsourcing complex projects that require significant expertise and are domain specific often does not save money. The cost of losing on-site personnel is also high. Hence, the decision to outsource should not be cost-motivated.

*D. Limitations*

According to Yin [40], interviews are one of the most important sources of case study information, and they should be considered "guided conversations rather than structured queries" [40]. The strength of holding interviews is that they are targeted, which means they are focused directly on case study topics. Interviews are also insightful and provide perceived causal inferences and explanations [40]. However, being aware of the weaknesses of interviews as evidence is vital. They are likely to be biased, both when it comes to poorly defined questions and responses. If an informant does not recall correctly, then his or her answers are inaccurate, and they can be reflexive in the way that the informant says what the interviewer wants to hear [40]. Formulating neutral, non-leading questions is essential. These were all things we were aware of when conducting and later analyzing the interviews.

VI. CONCLUSION AND FUTURE WORK

Developing software in teams distributed over continents is challenging, but agile software development and BizDevOps teams reduce the challenges of globally distributed teams. Although such approaches mitigate some of the challenges, the main reason companies terminate offshore contracts is the low quality of the software produced. Therefore, finding ways to improve quality is key to GSE. Currently, the use of PRs is a preferred way of reviewing code, mainly to improve the quality of the code and enable continuous deliveries.

We conducted an interpretative case study to investigate factors that affect the PR approach in distributed agile BizDevOps teams. We report on the friction caused by using PRs for distributed development as part of an offshore outsourcing relationship between two sites. We found multiple barriers to the code review process. Examples include off-site developers struggling with domain complexity, the use of two agile approaches (a flow-based Kanban approach and a timeboxed Scrum approach) in the distributed team leading to an overload of PRs in periods, and a lack of social networks between developers across sites (mainly because of high turnover off-site).

The developers studied on-site were frustrated by the tedious task of reviewing PRs from the offshore site. Future work should investigate ways to set up a successful PR approach that do not negatively affect the developers' motivation. Our findings showed the companies' KPIs did not measure the success of the distributed teamwork and the quality of the code produced. Governing and measuring in global software development is known to be difficult and research is lacking [41]. Our findings confirm previous research that showed the number of PRs might not be a good measurement for developer productivity [22].

ACKNOWLEDGMENTS

We are grateful to all the participants in the case study. The work was supported by the A-team project and the Research council of Norway through grant 267704.